\documentclass[reprint,amsmath,amssymb,aps,prb,groupedaddress,nofootinbib,twocolumn,superscriptaddress]{revtex4-2} % linenumbers
\usepackage{graphicx}
\usepackage{amsthm,amssymb,amsmath,braket,mathdots}
\usepackage{bm}
\usepackage[pagebackref=false,pdfnewwindow=true]{hyperref} %\hypersetup{draft}
\usepackage{epstopdf,psfrag}
\usepackage{relsize,amsbsy}
\usepackage[export]{adjustbox}

\usepackage{graphicx,xcolor} %tikz
\usepackage{tabularx, booktabs} %for tables with width

\usepackage{units} %\nicefrac,units
\usepackage{dsfont} %write unity 1

\newcommand{\ii}{{\mathrm{i}}} %imaginary unit is straight
\newcommand{\e}{{\mathrm{e}}} %Euler's number straight
\renewcommand{\v}[1]{\bm{#1}} %write vectors fat
\renewcommand{\vec}[1]{\bm{#1}} %write vectors fat

\renewcommand{\d}{\mathrm{d}}
\newcommand{\1}{\mathds{1}} %identity matrix

\DeclareMathOperator{\sinc}{sinc} %sinx/x

\renewcommand{\eqref}[1]{Eq.~(\ref{#1})}

\usepackage{tikz}
\usetikzlibrary{arrows.meta}
\definecolor{bananayellow}{rgb}{1.0, 0.88, 0.21}
\definecolor{straw}{rgb}{0.32, 0.28, 0.1}

\usepackage[caption=false]{subfig}
\usepackage{changes}
\definechangesauthor[name=Valentin, color=blue]{VL} 
\definechangesauthor[name=Johannes, color=red]{JK}

\begin{document}

\title{Numerical Study of Quantum Oscillations of the Quasiparticle Lifetime: \\ Impurity Spectroscopy, Novel Electric Field and Strain Effects}
%Numerical Study of Quantum Oscillations of the Quasiparticle Lifetime: Impurity Spectroscopy, Novel Electric Field and Strain Effects
%\title{A lattice approach to quasiparticle lifetime quantum oscillations}
\author{Valentin Leeb}
\affiliation{Technical University of Munich, TUM School of Natural Sciences, Physics Department, TQM, 85748 Garching, Germany}
\affiliation{Munich Center for Quantum Science and Technology (MCQST), Schellingstr. 4, 80799 M{\"u}nchen, Germany}
\author{Johannes Knolle}
\affiliation{Technical University of Munich, TUM School of Natural Sciences, Physics Department, TQM, 85748 Garching, Germany}
\affiliation{Munich Center for Quantum Science and Technology (MCQST), Schellingstr. 4, 80799 M{\"u}nchen, Germany}
\affiliation{Blackett Laboratory, Imperial College London, London SW7 2AZ, United Kingdom}
\date{\today}

\begin{abstract}
Quantum oscillation (QOs) measurements constitute one of the most powerful methods for determining the Fermi surface (FS) of metals, exploiting the famous Onsager relation between the FS area and the QO frequency. The recent observation of non-Onsager QOs with a frequency set by the difference of two FS orbits in a bulk three-dimensional metal can be understood as the QO of the quasiparticle lifetime (QPL) due to interorbital scattering [Huber, Leeb, {\it et al.}, Nature 621 (2023)]. QPL oscillations (QPLOs) generalize magneto-intersubband oscillations (MISOs) known from coupled two-dimensional metals. They may provide a novel tool for extracting otherwise hard-to-measure intra- versus interband scattering times of quasiparticles. 
Here, we provide a numerical lattice study of QPLOs comparing transport and thermodynamic observables. We explore the effect of different imperfections like general impurities, Hall effect-induced electric fields, various forms of strain from bending, and magnetic field inhomogeneities. We confirm the basic phenomenology of QPLOs as predicted in analytical calculations and identify additional novel, non-perturbative features. Remarkably, we find that some imperfections can stabilize, or even enhance, non-Onsager QPLOs in contrast to standard QO frequencies. We discuss various avenues for identifying QPLOs in experiments and how to use their dependence on imperfections to extract material properties.
\end{abstract}
	
\maketitle

\section{Introduction}
Quantum oscillation (QO) measurements are an exceptionally sensitive tool for measuring Fermi surface (FS) geometries as well as the strength of interaction effects via extracting the effective masses $m_i$ from the temperature dependence~\cite{deHaas1930,Onsager1952,Shoenberg}. Often, the observation of a QO frequency serves as a unique and sensitive indicator for the existence of a FS. For example, in underdoped cuprates, QO studies famously confirmed the presence of a closed FS pocket in a magnetic field~\cite{doiron2007quantum,sebastian2012towards}. Similarly, QO experiments in the spin density wave parent phase of iron-based superconducting compounds showed the emergence of small pockets, in contrast with angle-resolved photoemission spectroscopy (ARPES) results at the time~\cite{sebastian2008quantum,terashima2011complete,coldea2013iron}. Agreement on the electronic structure suggested by QOs was confirmed only much later~\cite{watson2019probing}.

The interpretation of QOs, as measured in transport or thermodynamic observables, is based on the famous Onsager relation, which ascribes each QO frequency to a semi-classical FS orbit~\cite{Onsager1952,Shoenberg}. Recently, unusual frequencies have been reported in the multi-fold semi-metal CoSi~\cite{huber2023quantum}, which are forbidden in the semi-classical Onsager theory and generalizations thereof~\cite{alexandradinata2023fermiology}. In addition, they are at odds with known mechanisms generating non-Onsager QO frequencies like magnetic breakdown, chemical potential oscillations, magnetic interaction, and (Stark) quantum interference~\cite{huber2023quantum,leeb2024a}. The emergence of these forbidden frequencies as combinations of two semiclassical QO frequencies, originating from underlying FSs, has been explained by a non-linear coupling of the FSs via interband impurity scattering~\cite{leeb2023theory}, dubbed quasiparticle lifetime oscillations (QPLOs) here. Ref.~\cite{huber2023quantum} argued that QPLOs are generic for bulk metals with multiple extremal orbits or multiple FS pockets. This claim is supported by the identification of many further candidate materials~\cite{leeb2024a}, ranging from topological semi-metals to iron-based superconducting parent compounds~\cite{leeb2024interband}. A key observation of Ref.~\cite{huber2023quantum,leeb2024interband} was that QPLOs could be erroneously interpreted as FS orbits, leading to incorrect identification of the electronic structure of a given material. 

The basic ingredient leading to QPLOs is a coupling between two QO orbits, such as interband impurity scattering. The inverse lifetime of the quasiparticles on a given orbit then is a sum of three terms -- a constant, the so-called Dingle temperature $T_{D1}$, an oscillatory contribution with frequency $F_1$ from intraorbit scattering, and, crucially, an oscillatory term with frequency $F_2$ determined by the other orbit 2 with an amplitude proportional to interorbit scattering. The constant Dingle temperature term damps the oscillations with a universal factor, known as the Dingle factor $R_D = \exp(-2\pi^2 m T_D/e B)$, with $e$ the electron charge and $B$ the magnetic field. The third term leads to cross terms of QOs in the conductivity with frequency $F_1$ and $F_2$ which appear as the sum and difference of the basis frequencies $F_1$ and $F_2$ in the conductivity
\begin{equation}
\sigma_{\pm} \propto \cos\left(2\pi \frac{F_1 \pm F_2}{B}\right) R_{D1} R_{D2} R_T\left(m_1 \pm m_2\right), 
\label{Eq1}
\end{equation}
where $R_T(m)$ is the Lifshitz--Kosevich dependence as introduced below in \eqref{eq:LK_dependence}. 

Conceptually, QPLOs generalize so-called magneto-intersubband oscillations (MISOs) known in coupled 2D electron gases~\cite{polyanovsky1988magnetointersubband} as well as quasi-2D materials \cite{polyanovsky1993hightemperature} and observed in several experiments \cite{leadley1989influence,coleridge1990intersubband,leadley1992intersubband,goran2009effect,mayer2016positive,abedi2020anomalous,minkov2020magnetointersubband}. MISOs are nowadays claimed to be understood in great detail \cite{raikh2023hightemperature} which is supported by several analytic works \cite{raikh1994magnetointersubband,averkiev2001theory,champel2002magnetic,grigoriev2003theory,thomas2008shubnikov,mogilyuk2018magnetic,raikh2023hightemperature}. The new insight of QPLOs is the generalization to bulk 3D materials~\cite{leeb2023theory} and a new qualitative picture of their origin~\cite{huber2023quantum}. Hereinafter, we refer to the phenomenon of additional QO frequencies, which originate in oscillatory contributions of the self-energy due to interorbit scattering, as QPLOs.

So far, all theories of QPLOs are based on analytical, perturbative calculations, requiring approximations like the self-consistent Born approximation for deriving \eqref{Eq1}~\cite{leeb2023theory} or even more ad-hoc assumptions. This calls for a controlled check of the analytic predictions, which numerical simulations can provide. The numerical evaluation of QOs in the presence of disorder is numerically not challenging, in the sense that the Hilbert space dimension grows only linearly with system size, which allows for an efficient numerical simulation at least for 2D systems, e.g., for transport \cite{leeb2024interband,krix2024quantum}. Besides supporting the phenomena itself, numerical simulations offer the possibility to test predictions of analytic theories in a more controlled manner than experiments can, provding additional insights to advance our understanding of QPLOs. In addition, one can capture effects beyond the limit of perturbation theory.  

In this work, we aim to benchmark analytical theories of QPLOs by full numerical lattice simulations of the Shubnikov--de Haas (SdH) effect, i.e., the conductance as function of the magnetic field. 
%The crystals, samples and experiments in which QOs are measured are never as perfect as assumed in theory. By now, we have a profound understanding how several types of imperfections affect QOs \cite{Shoenberg}, however these are all perturbative and never exact. 
In addition, we show how QPLOs respond to different types of impurities compared to standard Onsager QO frequencies, which paves the way for using QPLO measurements to understand otherwise hard to obtain impurity properties of materials. Beyond disorder, we also study the effects of i) electric fields from current-induced charge accumulation, ii) strain engineering from bending nanostructures, and iii) magnetic field inhomogeneities. The unique response of QPLOs to such imperfections may, on the one hand, serve as an additional criterion simplifying their identification and, on the other hand, allow to extract material properties.

The work is structured as follows. In sec.~\ref{sec:model} we introduce the lattice model and the numerical method.
In the main part, sec.~\ref{Sec:III}, we discuss the response of QPLOs to different types of impurities. Next, in sec.~\ref{sec:QPLO_analysis}, we analyze the properties of QPLOs like the appearance of higher harmonics, the dependence of QPLO on the non-linear coupling of the FSs, the phase relation to the basis frequencies and the behavior in thermodynamic observables. In the following sections, we discuss the effect of finite temperature, sec.~\ref{sec:finite temperature};  electric fields induced by a Hall effect from a varying current strength, sec.~\ref{sec:hall effect}); crystal strain induced by bending, sec.~\ref{sec:bending}; and inhomogeneities in the magnetic field, sec.~\ref{sec:magnetic field}.

% \begin{itemize}
%     \item introduce MISOs and differnece frequency oscillations in quasi-2D materials
%     \item subject to many theoretical analyitcal suties and experiments
%     \item claimed to be understood in great detail
%     \item recent interest under the name QO of QPLT, because mechanism seems to be more general and also observable in 3D
%     \item claim of Nature is that frequencies might be misinterpreted in several materials
%     \item impurity induced Difference frequency QO can be efficiently studied in the numerical context
%     \item offers the exploration of phenomena which are hard to evaluate analytically, example structural distortion like bending or an underlying Hall effect as background.
%     \item dependence on scattering beyond the self-consistent Born approximation
%     \item also checked well understood properties like finite temperature and higher harmonics.
% \end{itemize}

\section{Two-band Model and set-up}
\label{sec:model}
QPLOs appear in two and three-dimensions~\cite{leeb2023theory}. Here, we focus on a basic 2D two-band model that also captures the relevant physics of quasi-2D materials. The latter are characterized by weakly coupled layers and a nonzero interlayer tunneling leads to a warping of the FS such that 2 extremal orbits at $k_z=0$ and $k_z=\pi$ appear. To include these orbits determining the QOs, it is sufficient to model only 2 layers, which would for periodic boundary conditions in $z$-direction (which are trivial in that case) correspond to two discrete momenta.

\begin{figure}
    \centering
    \includegraphics[width=\columnwidth]{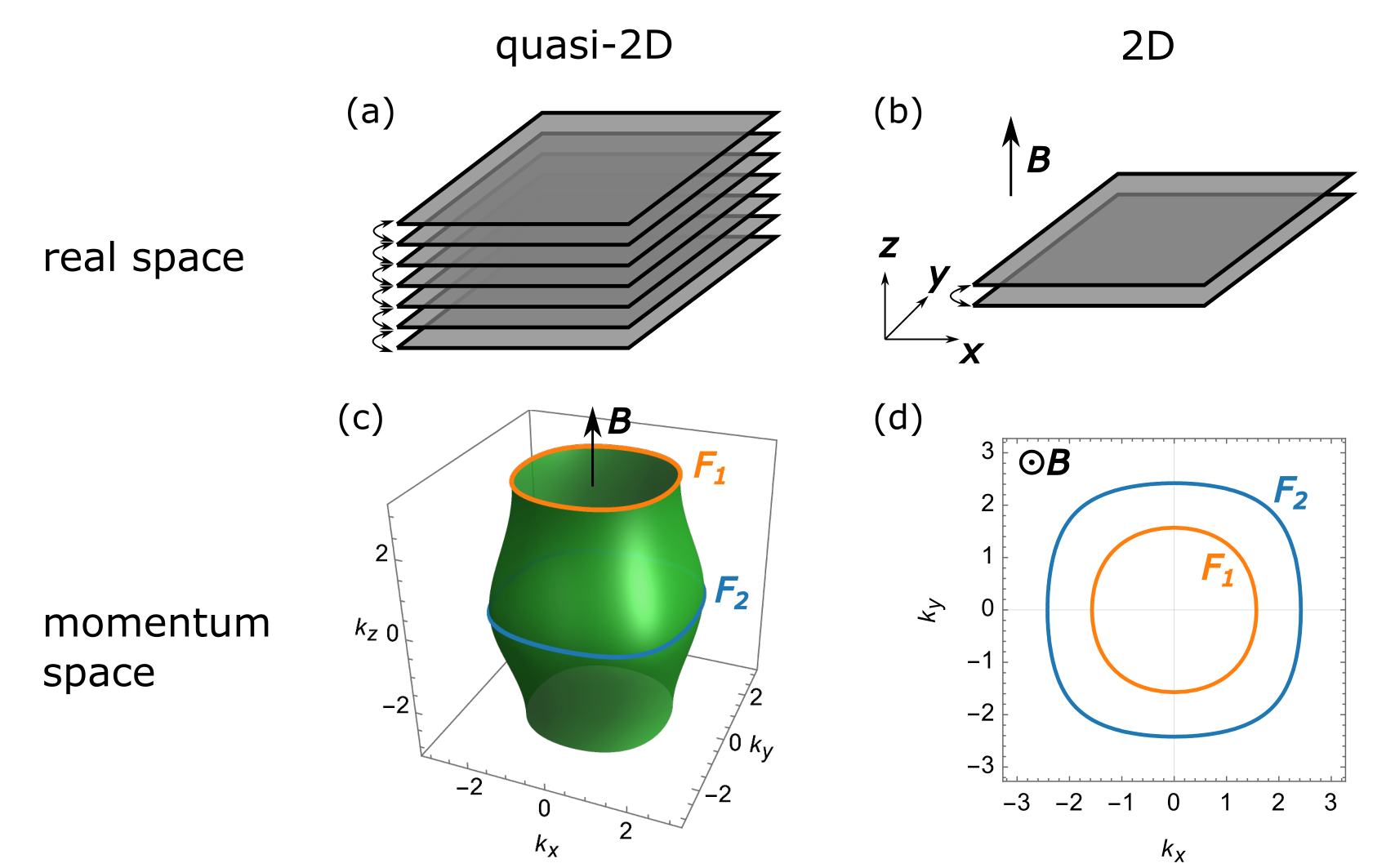}
    \caption{Reduction scheme to describe the extremal QO orbits of layered quasi-2D materials with a basic 2D  two-band model. (a) Quasi-2D materials consist of layers which are strongly coupled within each layer by $t, t'$ whereas the transfer integral in $z$-direction $t_\perp$ is smaller. (c) The FSs are typically warped cylinders where the extremal QO orbits for $\vec{B} \propto \hat z$ are located at $k_z = 0,\pi$. (d) An analogous description in 2D is just a simple two-band model where bands are split by $2 t_\perp$. (b) The corresponding real space picture is just a system with two layers where the bands correspond to the bonding and antibonding states of the layers.}
    \label{fig:fig1}
\end{figure}

\begin{figure*}
    \centering
    \includegraphics[width=\textwidth]{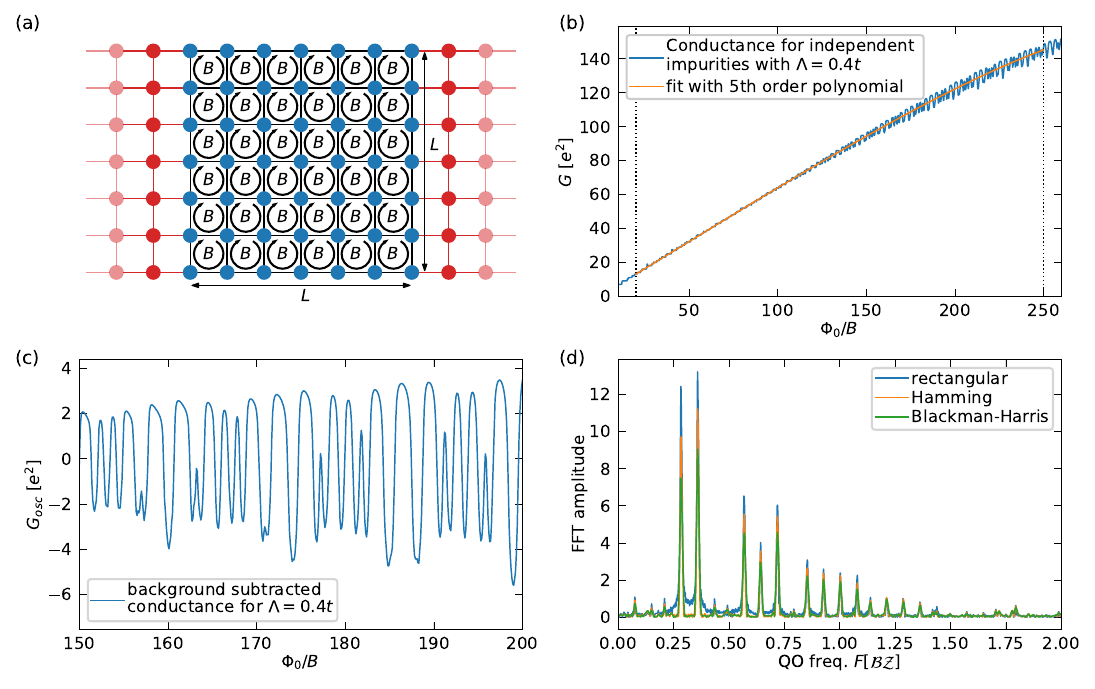}
    \caption{(a) shows a schematic sketch of the implemented model used for conductance simulations. The scattering region (blue vertices) of system size $L\times L$ implements the tight-binding model with impurities $H=H_0+H_{imp}$ in the presence of a magnetic field $B$. The leads (red vertices) are translational invariant, hence do not have impurities, and are not exposed to the magnetic field. The conductance $G$ from the left lead to the right lead is computed. (b) shows the exemplary result for the conductance $G$ for independent impurities. We fit the conductance in the field region $I_B = (20,250)$ (marked by the black dotted lines) with a $5^\text{th}$-order polynomial, which is then subtracted from the numerical data to obtain the oscillating part of the signal, see (c) for a part of the oscillating signal. To obtain a spectrum, (d) the oscillating signal is zero-padded, scaled with a windowinf function, and then Fourier transformed. The absolute of this, in general, complex-valued data is the fast-Fourier transformation (FFT) amplitude. The identified peaks are consistent for various window types.}
    \label{fig:analysis_procedure}
\end{figure*}

Hence, we consider the Hamiltonian
\begin{align}
H_\text{quasi-2D} =& -t\sum_{\langle i j \rangle, z} c^\dag_{i,z} c_{j,z} -t'\sum_{\langle\langle i j \rangle\rangle, z} c^\dag_{i,z} c_{j,z}
\nonumber \\
&-
\frac{t_\perp}{2} \sum_{\langle z, z' \rangle, i} c^\dag_{i,z} c_{i,z'}
\end{align}
on a square lattice, where $t$ ($t'$) are intralayer (next-) nearest-hoppings and $t_\perp \ll t$ is the interlayer hopping, see Fig.~\ref{fig:fig1}~(a) and (c). For simplicity we focus on \mbox{$z=0,1$} which is essentially the 2D limit, shown in Fig.~\ref{fig:fig1} (b) and (d), and introduce the  orbital (layer) basis $\Psi_i=(c_{i,0},c_{i,1})$ such that 
\begin{align}
H_0 =& -t\sum_{\langle i j \rangle} \v{\Psi}^\dag_{i} \v{\Psi}_{j}  -t' \sum_{\langle\langle i j \rangle\rangle} \v{\Psi}^\dag_{i} \v{\Psi}_{j}  - t_\perp \sum_{j} \v{\Psi}^\dag_{j} \tau^x \v{\Psi}_{j}\\
=& -\sum_{\vec{k}} \v{\Psi}^\dag_{\vec{k}} [2t( \cos k_x+\cos k_y) 
\nonumber \\
&
\qquad+ 4t' \cos k_x \cos k_y +t_\perp \tau^x ]\v{\Psi}_{\vec{k}}  \\
=& -\sum_{\vec{k}} E_\pm(\vec{k}) \v{\Phi}^\dag_{\vec{k}}\v{\Phi}_{\vec{k}} 
\end{align}
where we introduce the bands $E_\pm(\vec{k}) = -2t \cos k_x -2t \cos k_y -4t' \cos k_x \cos k_y \pm t_\perp$ and the fields in the band basis $\v{\Psi}_{\vec{k}} = U \v{\Phi}_{\vec{k}}$ which are just bonding and antibonding states 
\begin{equation}
    U = \frac{1}{\sqrt{2}}\begin{pmatrix} 1&1\\1&-1\end{pmatrix}.
\end{equation}

\subsection{Peierls substitution}
We apply an orbital magnetic field $B$ in the out of plane $z$-direction, by Peierls substitution, which transforms the hopping $t_{\vec{r},\vec{r}'}$ from site $\vec{r}$ to site $\vec{r}'$
\begin{equation}
t_{\vec{r},\vec{r}'} \rightarrow t_{\vec{r},\vec{r}'} \exp\left(-\ii e \int_{\vec{r}}^{\vec{r}'}\vec{A}(\vec{l}) \d \vec{l} \right).
\end{equation}
Here, $\vec{A} = (- y B,0,0)$ is the vector potential. For this gauge choice, only the hoppings $t_x$ in x-direction and the diagonal next-nearest neighbor hoppings $t_{xy}$ acquire a phase.

Additionally, we consider potential impurities
\begin{equation}
    H_{imp} = \sum_j \v{\Psi}^\dag_{j} \Lambda_j^\Psi \v{\Psi}_{j}
\end{equation}
where $\Lambda_i$ is, in general, a random hermitian matrix. In sec.~\ref{sec:impurity models} we specify different types of the scattering matrix $\Lambda_j^\Psi$ and study their effect on QOs and QPLOs.

\subsection{Numerical method}
\label{sec:numerical method}
Numerically, it is more convenient to work with dimensionless quantities. We set the lattice constant and $\hbar$ to 1. The dimensionless magnetic field $B/\Phi_0$ can be interpreted as the amount of flux through a single plaquette measured in units of the flux quantum $\Phi_0 = \frac{2\pi}{e}$.
%in python 2pi* B=(phi/2pi) (old plot)

Our goal is a full numerical lattice calculation of the SdH effect. This requires computing the conductance $G(B)$ of the model given by $H = H_0 + H_{imp}$ as function of the magnetic field $B$ and averaged over impurity samples. 

We evaluate the conductance through the built-in Landauer--B\"uttiker algorithm \cite{landauer1957spatial,landauer1970electrical,buttiker1986fourterminal,buttiker1988symmetry} of the python software package kwant \cite{kwant}. For this, we construct an open lattice, dubbed scattering region, of $L\times L$ sites with 2 orbitals per site and implement $H$ including the Peierls phases. Additionally, we attach two leads of width $L$ on the right and left side of the scattering region, see Fig.~\ref{fig:analysis_procedure}~(a). The leads realize $H_0$ without magnetic field and are characterized by translational invariance in $x$-direction. The Landauer--B\"uttiker algorithm evaluates the transmission probability $T_{nm}(E)$ from an eigenstate $n$ with energy $E$ of the left lead to an eigenstate $m$ with energy $E$ of the right lead via an $S$-matrix approach. The conductance is the sum over all channels at the Fermi energy $\mu$ \cite{landauer1957spatial,landauer1970electrical,buttiker1986fourterminal,buttiker1988symmetry}
\begin{equation}
    G = \frac{e^2}{2\pi} \sum_{n,m} T_{nm}(\mu).
\end{equation}

We evaluated the conductance $G$ numerically for 2500 equidistant values of $\nicefrac{\Phi_0}{B}$ from 10 to 250 and observed SdH oscillations as a function of $\nicefrac{\Phi_0}{B}$, see Fig.~\ref{fig:analysis_procedure}~(b). We then analyzed the Fourier transform in $\nicefrac{\Phi_0}{B}$ with standard QO techniques, which include subtraction of a 5th-order polynomial background to extract the oscillatory signal, shown in Fig.~\ref{fig:analysis_procedure}~(c), zero padding to increase the point density of the spectrum and windowing with a Blackman--Harris window to decrease spectral leakage. The representative steps are shown in Fig.~\ref{fig:analysis_procedure}. Fourier spectra, see Fig.~\ref{fig:analysis_procedure}~(d), are shown with frequencies in units of the BZ area. They show sharp peaks at the contributing QO frequencies.

We also implemented a numerical simulation of the de Haas--van Alphen (dHvA) effect. For this, we use the scattering region without additional leads. We then compute the density of states (DOS) $\rho(E)$ via the kernel polynomial method \cite{weisse2006kernel}, which is based on a Chebyshev expansion of the spectral density. For sampling, we used 30 randomly chosen vectors such that only the bulk of the system is sampled. Cutting off the edges suppresses finite-size boundary effects. We defined the bulk by the set of all lattice points at least 40 sites away from the edges and used 7000 Chebyshev moments.

The DOS can be used to determine thermodynamic observables. First, the zero temperature thermodynamic potential is computed 
\begin{equation}
\Omega(\mu) = \int_{-\infty}^\mu \d E \rho(E) ( E- \mu).
\label{eq:Omega0}
\end{equation}
The magnetization $M$ can then be computed via the derivative $M = -\partial \Omega/\partial B$. Here, we experienced that a numerical evaluation of the derivative of the data obtained with the kernel polynomial method is unstable. A different thermodynamic observable, the particle number $N$, is also defined via a derivative $N = -\partial \Omega(\mu)/\partial \mu$, however it can be carried out analytically to obtain 
\begin{equation}
N(\mu) = \int_{-\infty}^\mu \d E \rho(E).
\label{eq:particle_number}
\end{equation}
The particle number can hence be evaluated directly from $\rho$ without increasing the noise on the data by an additional numerical derivative.

The oscillating part of particle number and magnetization are related by $N \propto B \frac{m}{F} M$ for a single frequency $F$ with effective mass $m$. Hence, the main difference is that the QO amplitude of $N$ reduces for decreasing magnetic field $B$. This does not pose a problem in numerical simulations since the magnetic fields are relatively large. 

\section{Impurity spectroscopy with QPLO\lowercase{s}}
\label{Sec:III}
Within our basic model, different types of impurities can be classified according to their orbital/layer structure. We first discuss various types of impurity vertices and then show how they lead to qualitatively different effects. 

\subsection{Impurity models}
\label{sec:impurity models}
\subsubsection{Identical impurities}
In the simplest impurity model, the vertex at each point is taken to be proportional to the identity $\Lambda_i^\Psi = \lambda_i \tau^0$ and the prefactor is drawn from $\lambda_i \in [-\Lambda^0/2,\Lambda^0/2]$ randomly and uniformly. The physical picture is that each layer is an identical copy of the other layer, which may arise from systematic defects when growing the material or atoms sitting between the layers, hence having the same effect on both. A transformation to the band basis $\Lambda_i^\Phi = U^\dag \Lambda^\Psi_i U$ shows that the scattering vertex remains diagonal in the band basis, therefore leading only to pure {\it intraband} scattering.

\subsubsection{Opposite identical impurities}
The impurities at each lattice point are again perfectly correlated but with opposite signs in each layer, $\Lambda_i^\Psi = \lambda_i \tau^z$. Their strength is drawn from $\lambda_i \in [-\Lambda^z/2,\Lambda^z/2]$ randomly and uniformly. The corresponding physical picture could be dipoles located between the layers. A transformation to the band basis $\Lambda^\Phi = \lambda_i \tau^x$ shows that it gives rise to a pure {\it interband} coupling term.

\subsubsection{Independent impurities}
Both impurity models above are somewhat artificial. Therefore, we also consider impurities randomly distributed in each layer, i.e.,
\begin{align}
    \Lambda_i^\Psi &= \begin{pmatrix} \lambda_i &0\\0&\eta_i\end{pmatrix} 
\end{align}
where $\lambda_i,\eta_i \in [-\Lambda/2,\Lambda/2]$ are distributed randomly, uniformly, and independently. Crucially, the vertex is a superposition $\Lambda_i = (\lambda_i+\eta_i)/2 \tau^0 + (\lambda_i-\eta_i)/2 \tau^z$ of the identical and the opposite identical impurities. The probability distribution of $(\lambda_i \pm \eta_i)/2$ is the convolution of the 2 uniform (rectangular) probability distributions and hence a triangular probability distribution. Transforming this to the band basis shows that independent impurities in the layers lead to equally strong intra- and interband scattering, as expected from the form of $\Lambda_i^\Psi$.

\subsubsection{Random interlayer coupling}
Finally, there are interlayer impurities $\Lambda_i^\Psi = \lambda_i \tau^x$ or $\Lambda_i^\Psi = x_i \tau^y$
where $\lambda_i \in [-\Lambda/2,\Lambda/2]$ randomly and uniformly. Unlike the other disorder models, random interlayer coupling constitutes a form of bond disorder and can hence be complex. A varying distance between the layers will only influence the real part of the hoppings. A complex disordered part of the hopping may, for example, arise from locally trapped fluxes or random spin-orbit coupling (i.e., heavy impurities). 

Real interlayer impurities, i.e. $\Lambda_i^\Psi = \lambda_i \tau^x$, transform into the band basis as $\Lambda^\Phi \propto \tau^z$, hence, only leading to intraband contributions. In contrast, complex interlayer impurities, i.e., $\Lambda_i^\Psi = \lambda_i \tau^y$, contribute only to interband scattering because the scattering vertex transforms to $\Lambda^\Phi \propto \tau^y$. 

\begin{figure}
    \centering
    \includegraphics[width=\columnwidth]{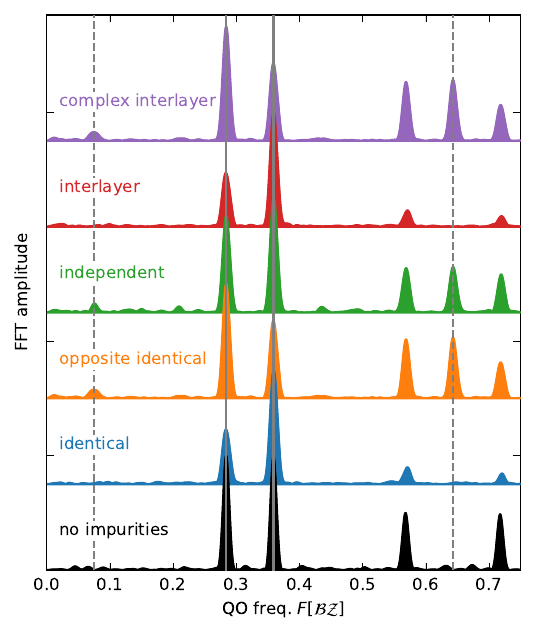}
    %\caption{The different discussed impurity models for impurity strengths $\Lambda/t=0,0.1,0.2$ (green, orange,blue). The main frequencies $F_1, F_2$ are marked by gray, solid lines and the combination frequencies $F_2\pm F_1$ by gray, dashed lines. Peaks at the combination frequencies only form for opposite identical and independent impurities. Additional peaks are higher harmonics. [ToDO: add plot for complex interlayer, 6th plot??]}
    \caption{QO spectra of the different impurity models for impurity strengths $\Lambda^\nu=0.4$ and for comparison in the absence of impurities (black). Spectra are normalized and shifted for clarity. The main frequencies $F_1, F_2$ are marked by gray, solid lines, and the combination frequencies $F_2\pm F_1$ by gray, dashed lines. QPLOs, visible by peaks at $F_2\pm F_1$, only appear for opposite identical, independent, and complex interlayer impurities. Additional peaks are higher harmonics. ($t_\perp = 0.4t$, $\mu=-0.5t$, $I_B = (20,250)$, Blackman--Harris window)}
    \label{fig:impurity models}
\end{figure}

\subsection{SdH oscillations}
Next, we show numerical results of the QO spectra for different impurity types.
QPLOs only emerge for an effective coupling of the semiclassical orbits~\cite{leeb2023theory}. In sec.~\ref{sec:impurity models}, we argue that in our model, only opposite identical and independent impurities lead to an effective interpocket coupling, whereas identical or interlayer impurities only contribute to intrapocket scattering processes. Therefore, we expect a strong dependence of the QO spectra on the types of impurities. 

We computed the SdH effect numerically for the different impurity models, following sec.~\ref{sec:numerical method}. Fig.~\ref{fig:impurity models} demonstrates that the main frequencies $F_1, F_2$ consistently appear for any model whereas the combination frequencies $F_2 \pm F_1$ only appear for sufficiently strong interpocket coupling, i.e., for opposite identical, independent and complex interlayer impurities. Indeed, the presence of a difference frequency peak from QPLOs may serve as tool discriminating between different forms of impurities. 

In the remainder of this work, we will focus on opposite identical and identical impurities, where the strength of interband scattering is $\Lambda^z$ and the strength of intraband scattering $\Lambda^0$.

\section{Analysis of QPLO\lowercase{s}}
\label{sec:QPLO_analysis}
\subsection{Higher harmonics}
QOs are not perfectly harmonic, i.e., their shape is not a perfect cosine. In the spectrum, this becomes visible by additional peaks at integer multiples $k F, k \in \mathbb{N}$ of the basis frequency $F$. The $k$th harmonic always appears together with $k^{th}$-order of the Dingle damping factor $R_D^k$ and is, therefore, typically decreasing in magnitude.

Research on QPLOs has so far focused on 2nd order frequencies, i.e., $F_2 \pm F_1$ which are of second order in the Dingle factor, namely governed by $R_{D1} R_{D2}$. Higher harmonics also exist for QPLOs as pointed out in Ref.~\cite{leeb2023theory}. For a model with two orbits, like our model here, higher harmonics of order $|k_1|+|k_2|$ can be found at $k_1 F_1 + k_2 F_2, k_i \in \mathbb{Z}$.

\begin{figure}
    \centering
    \includegraphics[width=\columnwidth]{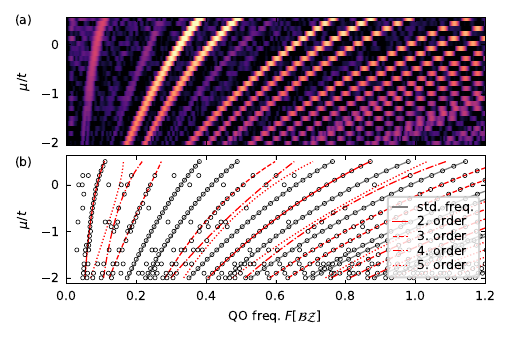}
    \caption{(a) FFT amplitude for various fillings $\mu$. (b) We extracted the position of all peaks above a detection limit defined by the 1.25 times the height of the low-frequency peak typically appearing in the data. The line indicate the analytic predictions for the basis frequencies and their higher harmonics (gray), and the combination frequencies (red) for different orders. ($t_\perp = 0.4t$, $\Lambda^z = 0.4 t$, $I_B = (50,250)$, Blackman--Harris window)}
    \label{fig:higher_harmonics}
\end{figure}

We analyzed the spectra for various different Fermi energies $\mu$. Fig.~\ref{fig:higher_harmonics}~(a) shows that a complicated QO spectrum with several peaks arises. We extract the peak positions of the QO spectrum and compare the results to analytic predictions, which we obtain from calculating the area of the semiclassical orbits and their integer and sum combinations. In panel (b), we demonstrate that nearly all peaks may be explained by higher harmonics of QPLOs up to 5th order following the analytic predictions~\cite{leeb2023theory}.

We note that for isolated parameters, additional peaks appear in the QO spectrum, which higher harmonics cannot explain. However, these do not seem to exist consistently for different Fermi energies. These unexplained peaks could, for example, be artifacts of the Fourier transformation, e.g., side lobes. We do not have an explanation for the peaks at low frequencies and high Fermi energies. Finally, we observe a weak splitting of some combination frequencies towards low Fermi energies, which we discuss in sec.~\ref{sec:magnetic field dependence}.

\subsection{dHvA oscillations}
The appearance of QOs in thermodynamic quantities, historically typically the magnetization/susceptibility, are known as the dHvA effect. dHvA oscillations are in many aspects similar to the ones appearing in the SdH effect and for standard Onsager QOs most aspects carry over. 

However, it turns out that QPLO are substantially different in transport compared to thermodynamic quantities. In analytic calculations for parabolic bands, all difference frequency combinations cancel exactly in the dHvA effect, whereas sum combinations do not~\cite{leeb2023theory}. This result also holds qualitatively true for relativistic (linear) dispersions, i.e., the amplitudes of difference frequency QOs are strongly suppressed. From the analytic theory, it is, therefore, unclear whether the absence of difference frequency QPLOs in the dHvA effect is generic. 

\begin{figure}
    \centering
    \includegraphics[width=\columnwidth]{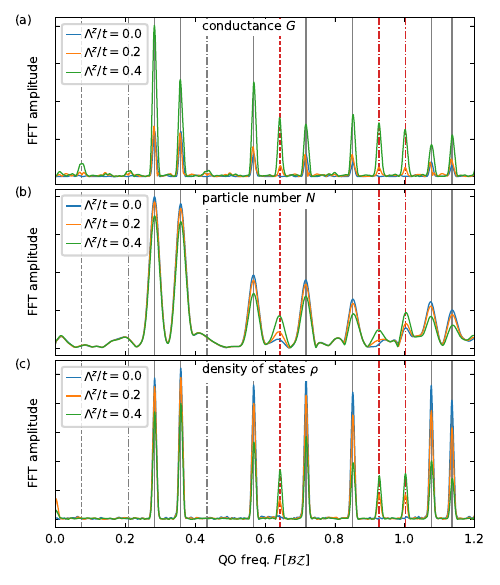}
    \caption{Comparison of the spectra of transport/ conductance, i.e., SdH effect (a) and thermodynamic quantities, i.e., dHvA effect, the particle number $N$ (b) and the DOS $\rho$ (c) as a function of increasing interband scattering rate $\Lambda/t$. The main frequencies $F_1, F_2$ and their higher harmonics $k F_1, k F_2$ are marked by gray, solid lines. Sum (Difference) frequency combinations are marked by red (gray) lines, dashed lines for second order, and dash-dotted lines for third order. Difference frequency combinations are only present in the conductance, i.e., in the SdH effect. ($t_\perp = 0.4t$, $\mu = -0.5 t$, $I_B = (20,250)$, Blackman--Harris window)}
    \label{fig:dHvA}
\end{figure}

In Fig.~\ref{fig:dHvA}, we compare quantities from the SdH effect (conductance) and from the dHvA effect (particle number and DOS) for increasing interband scattering rate. Fig.~\ref{fig:dHvA} confirms that difference frequency combinations of QPLOs are strongly suppressed in thermodynamic quantities for generic band structures, if not entirely absent. 

Our high-quality spectrum of the DOS, Fig.\ref{fig:dHvA}~(c), demonstrates that QPLOs cannot be understood on the basis of the DOS alone. Additional knowledge about the lifetime, hence the full spectral function, is required to explain them.

\begin{figure}
    \centering
    \includegraphics[width=\columnwidth]{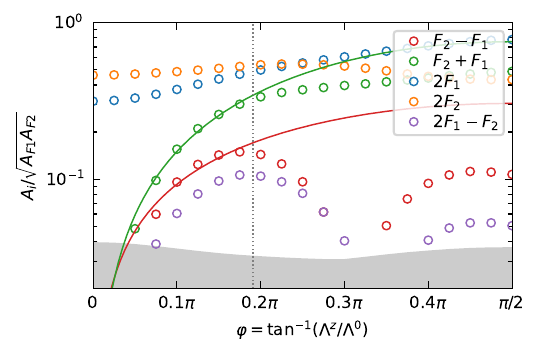}
    \caption{Relative amplitudes of main frequencies for different inter- vs. intraband scattering  ratios. $\Lambda^z = 0.4 t \sin \varphi$, $\Lambda^0 = 0.4 t \cos \varphi$ such that the Dingle temperature $\propto (\Lambda^0)^2 + (\Lambda^z)^2$ is constant. The solid lines indicate fits with $ \propto (\Lambda^z)^\alpha$, we find $\alpha_{F_2+F_1}=1.4\pm 0.1$ and $\alpha_{F_2-F_1}=1.0\pm0.2$. The gray shaded area indicates where amplitudes would be below the detection limit.}
    \label{fig:amplitude_relation}
\end{figure}

\subsection{Amplitude relation}
\label{sec:amplitude relation}
Next, we study the dependence of the QO amplitudes on the interband scattering $\Lambda^z$. In order to rule out the effect of a varying Dingle temperature, i.e., effects induced by the changing broadening of the LLs, we change the intra- $\Lambda^0$ and interband scattering $\Lambda^z$ such that their square sum $(\Lambda^0)^2 + (\Lambda^z)^2$, which is proportional to the Dingle temperature~\cite{leeb2023theory}, remains constant. 

Even though the Dingle temperature is kept constant, we find an increase of the main frequencies by roughly a factor of $8$ when comparing pure intraband coupling to pure interband coupling. Hence, we analyze the relative amplitudes of the higher harmonics with respect to the algebraic mean of the peak height of the two basis frequencies, see Fig.~\ref{fig:amplitude_relation}. The second harmonics $2F_1, 2F_2$ remain roughly constant as expected. The combination frequencies originating from QPLO vanish for decreasing interband coupling $\Lambda^z$. We fitted the amplitudes of the frequencies $F_2\pm F_1$ with $(\Lambda^z)^\alpha$ from 0 to 0.6 and found $\alpha \approx 1$. This result is at odds with analytic predictions. QPLOs arise in second order in the Born approximation and should, therefore, show a scaling behavior of $\alpha = 2$. The stronger scaling points towards the fact that QPLOs are either a lower order phenomenon, which does not hold in analytic calculations, or that higher order scattering contributes more strongly than expected. The stronger scaling indicates that the scattering expansion is non-perturbative, rendering the second-order approximation only qualitatively correct.

A second important observation from Fig.~\ref{fig:amplitude_relation} is the non-monotonic behavior of difference frequency combinations. The amplitudes of $F_2-F_1$ ($2F_1-F_2$) go to zero at around $\Lambda^z \approx 1.5 \Lambda^0$ ($\Lambda^z \approx 2 \Lambda^0$). We have checked that the zeros do not depend on the absolute strength of $\Lambda_z^2 + \Lambda_0^2$. It is unclear what the origin of these zeros is. 

%Higher order combination frequencies show a more complicated dependence on $\Lambda^z$. Probably due to the fact that higher order combination frequencies can appear as higher harmonics at second order scattering events, but also from higher order scattering events.

\subsection{Peak splitting}
\label{sec:magnetic field dependence}
A detailed analysis of the major peaks of the spectrum shows that, in some cases, they are actually split into two subpeaks, shown in detail in Fig.~\ref{fig:peak_splitting}~(a). The peak splitting is also visible in Fig.~\ref{fig:higher_harmonics}~(b) $F_2-F_1$ at low $\mu$, in Fig.~\ref{fig:dHvA}~(a) $F_2-F_1$ for $\Lambda/t=0.2$, in Fig.~\ref{fig:Hall_coefficient}~(a) $F_1$, $F_2$ at $e R_H I = 0.06 t$ and $F_1+F_2$ at $e R_H I = 0.03 t$. 

We show in the appendix sec.~\ref{app:peak_splitting} that peak splitting is independent of the impurity type. The physical origin is that due to the oscillating quasiparticle lifetime, the QO amplitudes are polynomial functions of the magnetic field \cite{leeb2023theory}. At the zeros of the amplitudes, the phase of the QOs jumps. Therefore, the phase itself becomes a function of the magnetic field. In the spectrum, a phase jump shows up as a weak peak splitting. However, peak splitting is of minor experimental relevance, because the zeros are located at values of $T_D/B$ which are mostly not tractable in experiment, see appendix sec.~\ref{app:peak_splitting}.

%We explain this behavior by additional contributions to the amplitude of each QO frequency by impurity scattering. The amplitudes are in fact polynomials in $B$ \cite{leeb2023theory}, hence the QO frequency may change phase as a function of $B$. In the spectrum this becomes visible by a weak splitting of the peaks. 

%The splitting of peaks is present in the numerical data for  high {\bf JK: Rally, not low? CHECK} magnetic fields. The zeros of the amplitudes typically appear for large magnetic fields, which means they are less relevant for experiments. Nevertheless, peak splitting may be an indication for the presence of strong interband scattering when large magnetic field ranges are scanned in experiment. 
%{\bf JK: If I rememeber correctly, even for F1=F2 you saw peak splitting in the main frequencies and a difference frequency appearing - correct? This would be interesting and we could add another figure?}

\subsection{Phase relation}
\begin{figure*}
    \centering
    \includegraphics[scale=1]{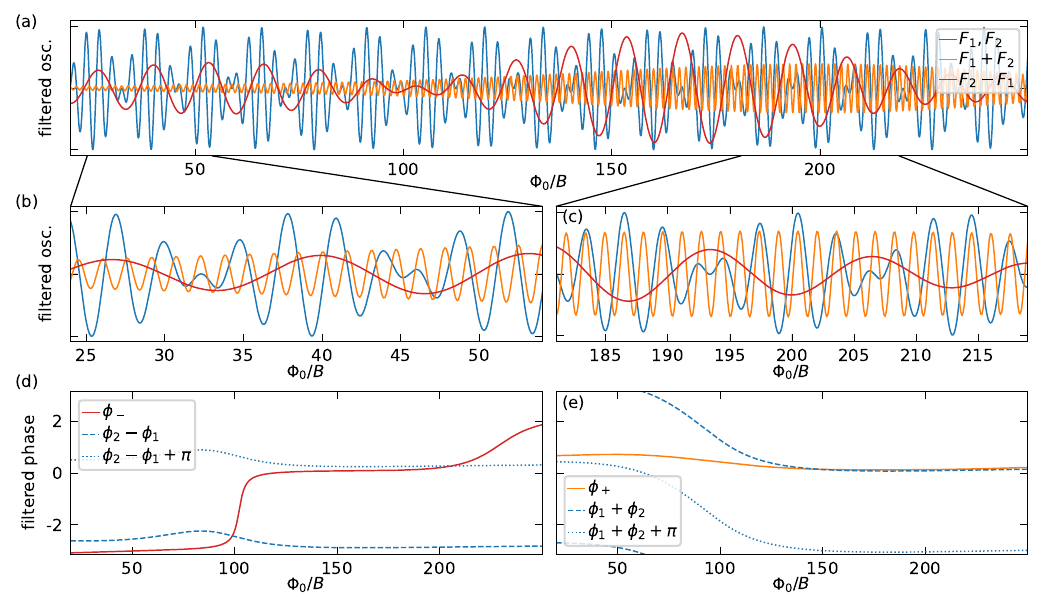}
    \caption{(a) Oscillations of the sum of the filtered basis frequencies (blue) show a beating pattern. The filtered oscillations of the difference frequency (red) envelope the beating pattern, indicating a fixed phase relation between the phase of the difference frequency $\phi_-$ and the difference of the phases of the main frequencies $\phi_2-\phi_1$. There is also a fixed phase relation for filtered oscillations of the sum frequency (orange) and the basis frequencies. At high (low) fields (b) ((c)), the minima (maxima) of the sum align with extrema of the beating pattern. (d) The filtered phase of the difference frequency (red) can be described by the difference of the filtered phases of the basis frequencies (blue). (e) The filtered phase of the sum frequency (orange) can be described by the sum of the filtered phases of the basis frequencies (blue). ($\Lambda^z=0.4t$, $\mu=-2t$)}
    \label{fig:phase_relation}
\end{figure*}
The phase of QOs is often of interest because of its relation to the Berry phase \cite{keller1958corrected,wilkinson1984examplea}. However, an absolute determination of the phase is usually difficult in experiment because large fields close to the quantum limit, i.e., $B \sim F$, of a QO frequency $F$ are required. In contrast, the relative phase between different frequencies can already be estimated by comparing the raw oscillatory data. 

Ref.~\cite{leeb2023theory,huber2023quantum} established, in contradiction with Ref.~\cite{grigoriev2003theory}, that there is a fixed phase relation between the phase $\phi_\pm$ of the QPLOs and the phases $\phi_i$ of the basis frequencies 
\begin{equation}
    \phi_\pm = \phi_2 \pm \phi_1 \quad\text{or}\quad \phi_\pm = \phi_2 \pm \phi_1 + \pi.
    \label{eq:phase_relation}
\end{equation}
The additional term $\pi$ accounts for the fact that the amplitudes may change sign, as also discussed in the appendix sec.~\ref{app:peak_splitting}.

We extracted the phases $\phi_i, \phi_\pm$ as function of the magnetic field by filtering, see Fig.~\ref{fig:phase_relation} (d,e). We also compared the relative phase of the, beating pattern, which is typical for close frequencies, with the filtered sum and difference frequencies, see Fig.~\ref{fig:phase_relation} (a-c). For a description of the filtering method, see appendix sec.~\ref{app:filtering}. We observe constant phases over large magnetic field ranges. $\phi_-$ experiences an abrupt phase change of $\pi$ at $\phi_0/B=100$. At the same magnetic field, $\phi_+$ shows a $\approx 0.16 \pi$ phase change.

Our analysis shows that the phase of QPLOs is indeed the sum/difference of the basis frequencies as described by \eqref{eq:phase_relation}. Our numerical results indicate that the even stronger relation 
\begin{align}
    &\phi_- = \phi_2 - \phi_1 \quad\text{and}\quad \phi_+ = \phi_2 + \phi_1 + \pi \quad \text{or}
    \nonumber \\
    &\phi_+ = \phi_2 + \phi_1 \quad\text{and}\quad \phi_- = \phi_2 - \phi_1 + \pi
    \label{eq:phase_relation2}
\end{align}
might hold, i.e., that the amplitudes of difference and sum frequency have always reversed signs. To our knowledge, this has not been discussed in the literature before. We conclude that a relative phase analysis, as done in Fig.~\ref{fig:phase_relation}, constitutes a strong indicator to confirm that a frequency in the spectrum can be assigned to QPLOs and not as a standard Onsager frequency.

\section{Finite temperature}
\label{sec:finite temperature}
The finite temperature conductance can be obtained within the Landauer-B\"uttiker formalism from  
\begin{equation}
    G_T = \frac{2 \pi}{e^2} \int_{-\infty}^\infty\d E  (-n_F'(E-\mu)) \sum_{n,m} T_{nm}(E),
    \label{eq:conductance finite temperature}
\end{equation}
a convolution with the derivative of the Fermi distribution function \cite{ryndyk2016theory}
\begin{equation}
    -n_F'(\epsilon) = \frac{\e^{\epsilon/T}}{T(1+\e^{\epsilon/T})^2}.
\end{equation}
For a numerical evaluation of this integral, we use that $n_F'(E-\mu)$ is strongly peaked around $\mu$ and limit the integration boundaries to $\approx \mu \pm 6T$ to capture more than 99.5\% of the spectral weight.

\begin{figure}
    \centering
    \includegraphics[width=\columnwidth]{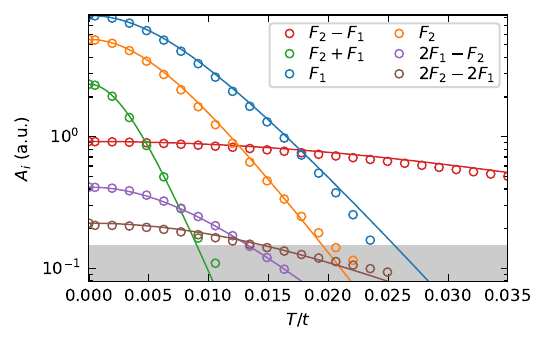}
    \caption{QO amplitudes of several different frequencies for different temperatures. The amplitudes are extracted from the peak maxima of the spectra at different temperatures. The solid lines show the expected LK dependence with $R_T(m)$ (\eqref{eq:LK_dependence}) where $m$ are combinations of $m_1 = 0.47/\Phi_0$ and $m_2 = 0.57/\Phi_0$ which are calculated from the band structure and for $\Phi_0/B$ we used the mean window point. ($t_\perp = 0.4 t, \mu=-t, \Lambda^z=0.4t$, $I_B=(100,250)$, Blackman--Harris window)}
    \label{fig:temperature_amplitude}
\end{figure}

We evaluated the zero temperature conductance $G(E)$ for several energies $E$ sampled logarithmically around the Fermi energy $\mu=-t$. Applying \eqref{eq:conductance finite temperature} we evaluated $G_T$ at $\mu=-t$ for several temperatures and then extracted the temperature dependence of the main peaks of the spectrum of $G_T$, see Fig.~\ref{fig:temperature_amplitude}.

Fig.~\ref{fig:temperature_amplitude} shows a perfect Lifshitz--Kosevich dependence with
\begin{equation}
    R_T(m) = \frac{\pi m T \Phi_0/B}{\sinh\left(\pi m T \Phi_0/B\right)}
    \label{eq:LK_dependence}
\end{equation}
of all QO frequencies. The effective mass \mbox{$m = e \d F(E)/\d E$} can be computed from the band structure by taking the derivative of the FS area with respect to the Fermi energy, obtaining $m_1 = 0.47/\Phi_0$ and $m_2 = 0.57/\Phi_0$. The theory of QPLO predicts for combination frequencies $k_1 F_1 + k_2 F_2$ a Lifshitz--Kosevich temperature dependence with an effective mass $k_1 m_1 + k_2 m_2$. The numerical data in Fig.~\ref{fig:temperature_amplitude} confirm this result with high precision. Most importantly, it confirms the presence of the nearly temperature-independent difference frequency for $m_1 \approx m_2$. 

We note that small deviations between the numerical results and the expected theory curve appear for increasing temperature. The deviations increase with decreasing mean of $I_B$, i.e, with increasing magnetic field. We suggest that this is due to our method of extracting the temperature dependence from the Fourier transform, which is to some extent inaccurate because of the dependence of $R_T(m)$ on the magnetic field $B$. We verified that higher order corrections to the Lifshitz--Kosevich factor are much smaller and are therefore not responsible for discrepancies between numerics and analytics.~\footnote{Lifshitz--Kosevich dependence assumes that the frequency can be linearly expanded around the chemical potential $F(\mu+\epsilon) = F(\mu) + \epsilon F'(\mu)$ where $F'(\mu)$ determines the effective mass and that higher order corrections are sufficiently small.}

\section{Electric Field Effects}
\label{sec:hall effect}
Next, we explore the effect of different types of additional perturbations on the QO spectrum. 

In a typical magneto-transport set-up to measure the SdH effect, a current $I$ is sent through the sample to measure the resistivity. Naturally, the set-up is susceptible to a classical Hall effect, where charge accumulates on the edges parallel to the current directions. The strength of the Hall effect can be characterized by a sample-dependent Hall coefficient $R_H$. The generated perpendicular Hall voltage $U_H = R_H I B/\Phi_0$ leads to an electric field affecting the QO spectrum.

We model the effect of a Hall voltage-induced electric field in our system by a spatially varying Fermi energy $\mu \rightarrow \mu(y) = \mu + e U_H y/L = e R_H I B y/L \Phi_0$. The leads are kept at constant Fermi energy.

\begin{figure}
    \centering
    \includegraphics[width=\columnwidth]{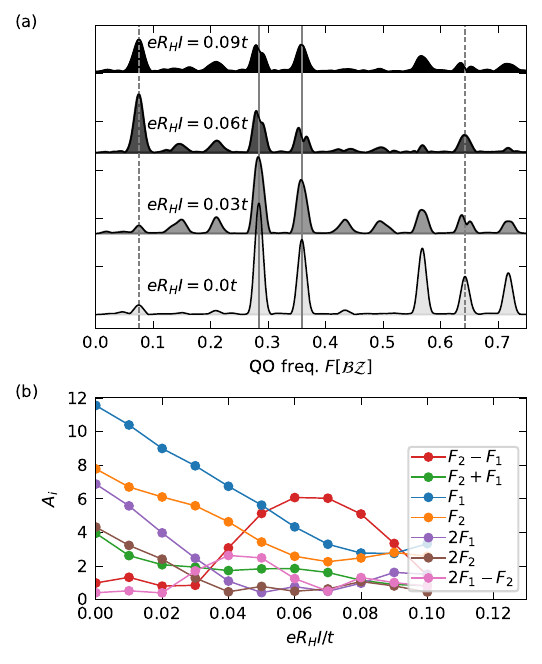}
    \caption{QO spectrum in the presence of finite electric field induced by a classical Hall effect. (a) Exemplary QO spectra for various currents $e R_H I$. The spectra are offset for clarity. The main frequencies $F_1, F_2$ are marked by gray, solid lines and the combination frequencies $F_2\pm F_1$ by gray, dashed lines. (b) Peak maxima of various frequencies as function of the current/Hall coefficient $e R_H I$. ($t_\perp= \Lambda^z = 0.4 t, \mu=-0.5t$, $I_B=(80,250)$, Blackman--Harris window)}
    \label{fig:Hall_coefficient}
\end{figure}

A qualitative picture of the effect of a spatially varying Fermi energy can be obtained by a phase smearing argument \cite{Shoenberg}. Note that formally, the application of a phase smearing argument requires that variations appear only on a length scale longer than the size of a cyclotron orbit. Here, $\mu(y)$ varies continuously within the cyclotron orbit. According to phase smearing, the damping of the QO frequencies can be evaluated by averaging over all present Fermi energies
\begin{align}
&\frac{1}{L} \int_{-L/2}^{L/2} \d y \cos \left(2\pi \frac{F(\mu+e U_H y/L)}{B}\right)
\nonumber \\
\approx& \int_{-1/2}^{1/2} \d u \cos \left(2\pi\frac{F +m U_H u}{B}\right)
\nonumber \\
=& \cos \left(2\pi\frac{F}{B}\right) \sinc\left(\pi \frac{m R_H I}{\Phi_0}\right)
\label{eq:hall_damping}
\end{align}
%\sinc(\pi m U_H/B) = \sinc(\pi m R_H I/\Phi_0) = \sinc(e m R_H I/2)
where $\sinc x = \sin x/x$.

The numerical results are shown in~Fig.~\ref{fig:Hall_coefficient}.
We observe a stronger decay of the main frequencies with respect to $e R_H I$ compared to the analytical prediction in Eq.~\eqref{eq:hall_damping}. We attribute this to the fact that the cyclotron orbits are not sufficiently large compared to the variation of the Fermi energy. Hence, the assumptions for the derivation of \eqref{eq:hall_damping} are not met. Nevertheless, the derivation provides an intuitive picture of the qualitative effect. With increasing electric field (or Hall coefficient/current), the QO amplitude should monotonically decrease to zero with a scale which is determined by the effective mass of the QO frequency. If this semiclassical phase smearing picture can be applied similarly to finite temperature, it implies that the amplitude of the difference frequency decreases weakly with $m_2-m_1$.

The main finding of this section is the anomalous behavior of the difference frequency combinations $F_2-F_1$ and $2F_1-F_2$ in Fig.~\ref{fig:Hall_coefficient}. Both amplitudes show a significant non-monotonic increase of more than a factor 6 before dropping again. This behavior is at odds with the present understanding of QPLOs or QOs in general.

\section{Strain Effects from Bending}
\label{sec:bending}
Intense material research in recent years has led to new developments, which allow strain tuning of materials in a controlled manner~\cite{hicks2014piezoelectric}. Specifically, there exist new experimental set-ups for continuously bending materials~\cite{diaz2021bending,MollPrivate}. 

\begin{figure}
    \centering
    \includegraphics[width=\columnwidth]{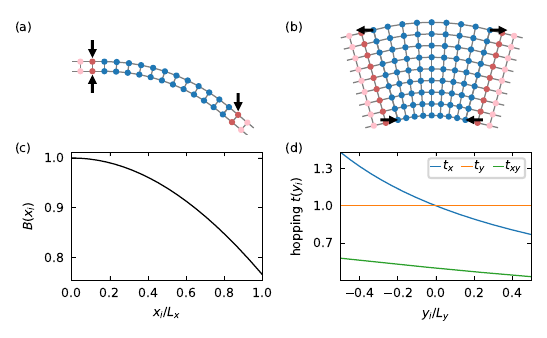}
    \caption{Different forms of bending: (a) Bending around the $y$-axis, here $\kappa_y=0.7$, leads to a decrease of the flux piercing through a single plaquette, see panel (c). (b) Bending around the $z$-axis, here $\kappa_z=0.6$ varies the longitudinal distance $a$ between the atomic orbitals. We assume $t \propto 1/a$ and obtain a spatial dependence of the longitudinal hoppings ($t_x$) and the next-nearest neighbor hoppings ($t_{xy}$), see panel (d).}
    \label{fig:bending_explain}
\end{figure}

Bending may have different effects depending on the microscopic details of the set-up and material. However, two effects appear quite generically and are straightforward to incorporate in our numerical implementation: (i) Bending leads effectively to a spatially dependent magnetic field by deforming the area of every single plaquette of the crystal differently, which changes the flux piercing through a plaquette. This effect becomes dominant when bending around an axis perpendicular to the magnetic field, e.g., the $y$-axis, see Fig~\ref{fig:bending_explain}~(a) and (c). (ii) The distance between the atoms in the crystal and, hence, the overlap of the atomic orbitals changes. This leads to spatially dependent hoppings, see Fig.~\ref{fig:bending_explain}~(b) and (d). This effect becomes dominant when bending in the plane perpendicular to the magnetic field.

Note that bending inside a material cannot only arise from external application of forces. Crystal imperfections may lead to similar phenomena, which become apparent in domains of slightly different material parameters and grain boundaries.

Here, we consider the effects of bending types (i), (ii) separately in order to understand their effects on QOs in a controlled manner. For simplicity, we focus on lattices with uniform curvature. We quantify the bending strength by the curvature $\kappa_j$ around the axis-$j$ in the middle of the lattice. The leads are kept at zero curvature.  

\subsection{Bending-induced effective magnetic field}
A non-zero value of $\kappa_y$ leads to a spatially dependent magnetic field $B(x) = B(0) \cos(\kappa_y x/L_x)$ through each plaquette, where $B(0) = B$ is the flux per plaquette of the relaxed model, see Fig.~\ref{fig:bending_explain}~(a) and (c). Note that the spatial dependence of the magnetic field is weak on the scale of the lattice, hence the vector potential $\vec{A}$ inherits the spatial dependence of the magnetic field. 

The naive expectation is that the electrons only feel the average magnetic field acting on the system $\Bar{B} = \int_0^{1} B(x) \d (x/L) = B \sinc \kappa_y$ as it has already been discussed in the standard literature \cite{Shoenberg}. Better estimates can be made by employing a phase smearing argument as \eqref{eq:hall_damping}. However, even for $\kappa_y \ll 1$ only small analytic progress can be made favoring a qualitative picture. The varying magnetic field damps the oscillations. Higher frequencies are damped stronger than lower frequencies.
% \begin{align}
%     &\int_{0}^{1} \d u \cos \left(2\pi \frac{F}{B \cos(\kappa_y u)}\right)
% \nonumber \\
% \approx& \int_{0}^{1} \d u \cos \left(2\pi \frac{F}{B} \left[1+\frac{(\kappa_y u)^2}{2}\right]\right)
% \nonumber \\
% =& \cos \left(2\pi \frac{F}{B} \right) \int_{0}^{1} \d u \cos \left(2\pi \frac{F}{B} \frac{(\kappa_y u)^2}{2}\right)
% \nonumber \\
% &- \sin \left(2\pi \frac{F}{B} \right) \int_{0}^{1} \d u \sin \left(2\pi \frac{F}{B} \frac{(\kappa_y u)^2}{2}\right)
% \end{align}

\begin{figure}
    \centering
    \includegraphics[width=\columnwidth]{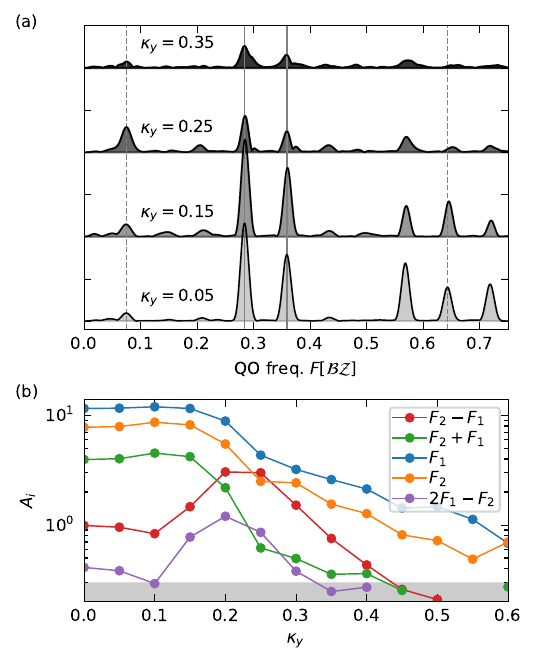}
    \caption{QO spectrum in the presence of finite bending in $y$-direction, i.e., spatially dependent magnetic field. (a) Exemplary QO spectra for various bendings $\kappa_y$. The spectra are offset for clarity. The main frequencies $F_1, F_2$ are marked by gray, solid lines and the combination frequencies $F_2\pm F_1$ by gray, dashed lines. (b) Peak maxima of various frequencies as function of the curvature $\kappa_y$ in $y$-direction. ($t_\perp= \Lambda^z = 0.4 t, \mu=-0.5t$, $I_B=(80,250)$, Blackman--Harris window)}
    \label{fig:bending_kappa_y}
\end{figure}

Our numerical observations contradict this naive expectations, see Fig.~\ref{fig:bending_kappa_y}. At first, the amplitudes of semiclassical frequencies and sum combinations of QPLO remain relatively stable with increasing curvature $\kappa_y$ before eventually decreasing. Most remarkably is the behavior of the amplitudes of the difference frequency combinations $F_2-F_1$ and $2F_1-F_2$. The amplitudes first decrease until $\kappa_y=0.1$ but then increase by roughly a factor of 3.

\subsection{Spatially varying hoppings}
Next, we study spatially dependent hoppings, see Fig.~\ref{fig:bending_explain}~(b) and (d). Similar to above, we expect that bending leads to a monotonic decrease of all QO amplitudes. 

\begin{figure}
    \centering
    \includegraphics[width=\columnwidth]{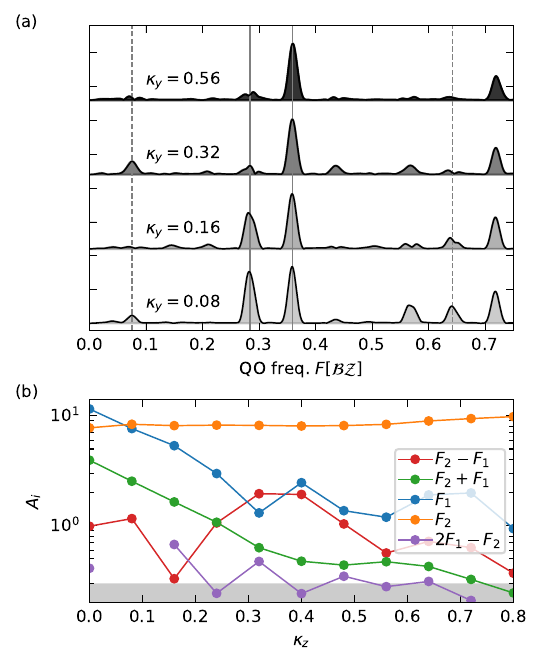}
    \caption{QO spectrum in the presence of finite bending in $z$-direction, i.e., spatially varying hoppings. (a) Exemplary QO spectra for various bendings $\kappa_z$. The spectra are offset for clarity. The main frequencies $F_1, F_2$ are marked by gray, solid lines and the combination frequencies $F_2\pm F_1$ by gray, dashed lines. (b) Peak maxima of various frequencies as function of the curvature $\kappa_z$ in $z$-direction. ($t_\perp= \Lambda^z = 0.4 t, \mu=-0.5t$, $I_B=(80,250)$, Blackman--Harris window)}
    \label{fig:bending_kappa_z}
\end{figure}

We observe a strong, almost monotonic decrease of the QO amplitude of $F_1$, whereas the amplitude of $F_2$ remains stable with the tendency to even increase (factor $\approx 1.3$), see Fig.~\ref{fig:bending_kappa_z}. Similar to above, the difference frequency increases in amplitude by a maximum factor of roughly $2$, peaking around $\kappa_z \approx 0.35$. 

We have verified that the stability of $F_2$ is also present without interband coupling. The stability probably arises due to the quantum Hall limit where transport is mediated by edge states and is not generic for QOs.

\section{Magnetic field inhomogeneities}
\label{sec:magnetic field}
Finally, we study the effect of inhomogeneous magnetic fields. Historically, unwanted field inhomogeneity posed a technical challenge in the early days of QOs, as already noted in Landau's seminal work with the original prediction of QOs ~\cite{landau1930diamagnetismus}. We explore the possibility of magnetic fields which are not constant over the sample. A possible reason for this is a spatially dependent magnetic susceptibility $\chi(x,y)$, which changes the magnetic field $B \rightarrow B(1+\chi(x,y))$. For simplicity we model $\chi(x,y) = \chi_0 (\sin(2\pi x/L_x)+\sin(2 \pi y/L_y))$.

An increasing magnetic field inhomogeneity should decrease the QO amplitude, as already discussed in the literature \cite{Shoenberg}. The limit at which the damping becomes relevant is $\chi_0  \approx B/\pi F$. From this condition, it is apparent that higher frequencies are affected more strongly or earlier than lower frequencies.

\begin{figure}
    \centering
    \includegraphics[width=\columnwidth]{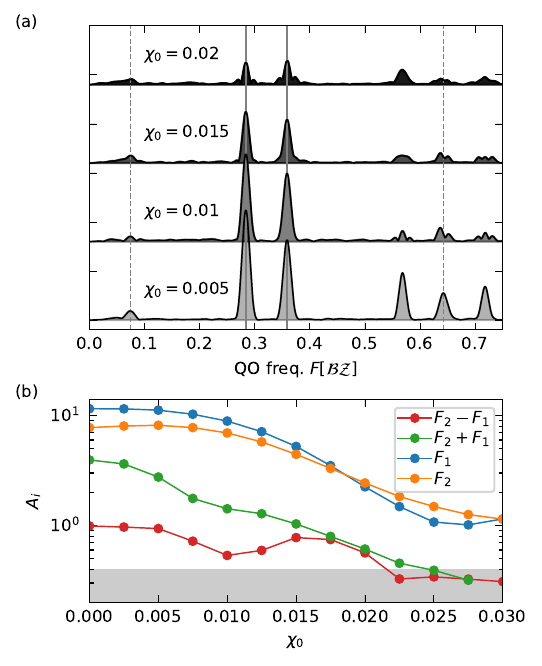}
    \caption{QO spectrum in the presence of magnetic field inhomogeneities, i.e., $B \rightarrow B(1+\chi(x,y))$. (a) Exemplary QO spectra for various strengths of the field inhomogeneity $\chi_0$. The spectra are offset for clarity. The main frequencies $F_1, F_2$ are marked by gray, solid lines and the combination frequencies $F_2\pm F_1$ by gray, dashed lines. (b) Peak maxima of various frequencies as function of the magnetic field inhomogeneity $\chi_0$. ($t_\perp= \Lambda^z = 0.4 t, \mu=-0.5t$, $I_B=(80,250)$, Blackman--Harris window)}
    \label{fig:chi}
\end{figure}

Fig.~\ref{fig:chi} shows the QO spectrum and the amplitudes for various $\chi_0$. As expected, the QO amplitudes decrease monotonically. However, yet again, we find an unusual non-monotonic behavior for the difference frequency $F_2-F_1$. After an initial decay, it peaks at roughly the same value as for zero inhomogeneity before decaying below the detection limit. 

\section{Summary and discussion}
In this work, we studied SdH and dHvA QOs in the presence of different types of impurities, electric fields, bending-induced strain, and magnetic field inhomogeneities. The motivation for this quantitative study comes from the analytical  prediction~\cite{leeb2023theory} and observation~\cite{huber2023quantum} of a new mechanism of non-Onsager difference frequency QO dubbed QPLOs, which we benchmarked here in a full numerical lattice implementation. 

We focused on a minimal but generic model featuring two FSs leading to two QO frequencies $F_1$ and $F_2$. Different types of impurities are introduced and by transforming the impurity vertex into the band basis, we evaluated the strength of intra- and interband coupling between the two semiclassical QO orbits $F_1$, $F_2$. We confirmed numerically that new QPLO frequencies at $F_2 \pm F_1$ appear only in the presence of nonzero interband coupling. We also observed higher harmonics $k_1 F_1 + k_2 F_2$ of the QPLO up to 5th order. We confirmed that the temperature dependence of $k_1 F_1 + k_2 F_2$ is an exact Lifshitz--Kosevich with an effective $k_1 m_1 + k_2 m_2$. In particular, we observed the temperature stability of $F_2-F_1$, making it the only observable frequency for $T>0.025t$, a tell-tale prediction for their experimental identification. 

In accordance with the theory for parabolic and linear bands \cite{leeb2023theory}, all difference frequency contributions $F_2-F_1$, $2 F_2 - F_1$, ... vanish in the dHvA effect. For parabolic bands, the cancellation of difference frequency contributions in analytics appears somewhat fine-tuned. Therefore, the exact absence in numerics points towards a more general reason that transport and thermodynamics quantities behave qualitatively differently for QPLOs.

In sec.~\ref{sec:amplitude relation}, we showed that the amplitudes of \mbox{QPLOs} scale linearly with interband coupling and not as expected quadratically. This shows that the expansion of the self-energy in the self-consistent Born approximation is non-perturbative, rendering the approximation only qualitatively correct. Additionally, we observe that difference frequencies may cancel even for finite interband couplings for particular parameter values.

In the remainder of our work, we showed that the relative amplitude of QPLOs increases, i.e., it has a qualitatively different behavior from the main Onsager frequencies under several different types of imperfections. The imperfections include electric fields induced by currents through the sample, i.e., a finite Hall effect, bending of the sample such that hoppings or the magnetic field acquire a spatial dependence and fluctuations of the magnetic field. In the case of the Hall effect and spatially fluctuating magnetic field, we find that the absolute amplitude of $F_2-F_1$ increases substantially. The amplitude dependencies are at odds with existing analytical theories, i.e., they cannot be explained by simple phase smearing arguments, as for finite temperature, or by a combination of the amplitudes of the basis frequencies. We verified that none of the above imperfections induces combination frequencies on its own. The unique response of the frequency $F_2-F_1$ to these imperfections demonstrates the non-perturbative nature of this QO frequency arising from intriguing interference effects not captured by semi-classical arguments. 

In the absence of interband scattering, we never observe combination frequencies in our exact numerics, even for non-trivial imperfections. Thus, interband scattering is key for inducing these new non-Onsager frequencies. In turn, an observation thereof can be used to perform 'impurity spectroscopy,' e.g., extracting otherwise hard-to-measure properties of intra- versus interband/orbital impurity contributions~\cite{leeb2024interband}.

Our work has established rigorously the key aspects of difference frequency QPLOs and unearthed a whole range of new phenomena beyond the perturbative theory of Ref.~\cite{leeb2023theory} pointing to effects beyond the lowest order Born approximation of interband scattering, which will be very worthwhile to explore in the future. In addition, interband scattering may also arise naturally from interactions and collective bosonic excitations like phonons or (para-)magnons. While similar QPLOs are expected in these cases neither analytical nor numerical works exist and are an important, yet challenging, avenue for future research. Again, the hope is that QPLOs may serve as a novel tool for extracting otherwise hard-to-obtain material properties.

{\it Data and code availability.--}
Code and data related to this paper are available on Zenodo \cite{code} from the authors
upon reasonable request.

\begin{acknowledgments}
We acknowledge helpful discussions and related collaborations with  N.~Huber, P.~Bieniek, M.~A.~Wilde, and C.~Pfleiderer. We would like to thank P Moll for pointing us to the idea of studying QO in strained materials. 
 V.~L. acknowledges support from the Studienstiftung des deutschen Volkes. 
We acknowledge support from the Imperial-TUM flagship partnership. JK acknowledges support from the Deutsche Forschungsgemeinschaft (DFG, German Research Foundation) under Germany’s Excellence Strategy–EXC–2111–390814868, DFG grants No. KN1254/1-2, KN1254/2- 1, and TRR 360 - 492547816; as well as the Munich Quantum Valley, which is supported by the Bavarian state government with funds from the Hightech Agenda Bayern Plus.
 \end{acknowledgments}

\clearpage
\bibliography{bib,zotero_bib}
\clearpage
\appendix
\renewcommand\thefigure{\thesection.\arabic{figure}} %new figure labels
\setcounter{figure}{0}

\section{Filtering}
\label{app:filtering}
\begin{figure*}
    \centering
    \includegraphics[width=\textwidth]{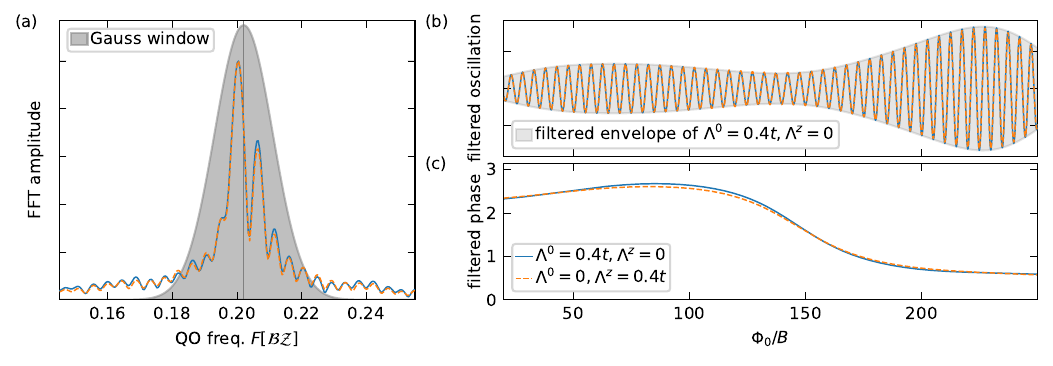}
    \caption{Phase dependence of the filtered main frequencies and filtering method. (a) Zoom-in of the SdH QO spectra to the first harmonics $F_1=F_2$ (gray solid line) for $t_\perp = 0$ ($\mu=-2t$, $I_B=(20,250)$, rectangular window). The peak is split into two subpeaks. The spectrum is scaled with the Gauss window (gray area) before performing an inverse Fourier transformation. (b) The absolute of the inverse Fourier transformation signal is the filtered signal. Computing the inverse Fourier transformation of the shifted, windowed spectrum yields the filtered envelope, the shaded area in (b), and the filtered phase, shown in (c).}
    \label{fig:peak_splitting}
\end{figure*}

For the filtering process, we first compute a complex-valued spectrum $s(F)$ with a rectangular window, as described in Fig.~\ref{fig:analysis_procedure}. To filter a frequency $f$, we scale the spectrum $s(F)$ with a narrow Gaussian window $\exp(-(F-f)^2/\Delta f^2)$ centered around $f$, see Fig.~\ref{fig:peak_splitting}~(a), and compute the inverse Fourier transform of $\exp(-(F-f)^2/\Delta f^2) \times s(F)$. The real part of the inverse Fourier transform is the filtered oscillation, see Fig.~\ref{fig:peak_splitting}~(b). The absolute (complex phase) of the inverse Fourier transform of $\exp(-(F-f)^2/\Delta f^2) \times s(F-f)$ are the filtered envelope (phase) of the filtered oscillation, see Fig.~\ref{fig:peak_splitting}~(b) (Fig.~\ref{fig:peak_splitting}~(c)).

\section{Peak splitting}
\label{app:peak_splitting}
We filter the frequencies as described in the appendix sec.~\ref{app:filtering}. In Fig.~\ref{fig:peak_splitting} we show that in the case of $t_\perp=0$, i.e., the case of a single degenerate QO frequency, there is a remarkable agreement of the conductance for interband and intraband scattering. Especially, peak splitting is not a result of interband scattering but of scattering in general, as expected from analytics \cite{leeb2023theory}. For inter- and intraband scattering, a phase change around $\Phi_0/B=150$ is the origin of the peak splitting. The phase change is continuous and roughly $0.6\pi$. It is unclear why the phase change differs from $\pi$.

The magnetic field at which the phase jump occurs increases as scattering, i.e., $T_D$ increases. This is expected from the zeros of the polynomial amplitudes. E.g. for the first harmonic, the zeros are at $T_D \approx e B/2\pi m$. Inserting the zero in $R_D \approx 0.04$ shows that the Dingle factor strongly damps the amplitude of QOs around the zero.

\begin{figure}
    \centering
    \includegraphics[width=\columnwidth]{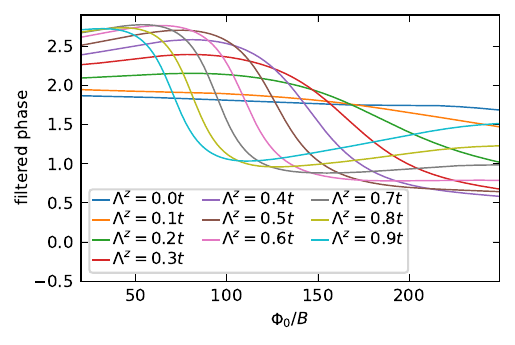}
    \caption{The filtered phase as function of magnetic field for degenerate frequencies, i.e., $t_\perp=0$ ($\mu=-2t$), and various values of interband scattering. Increasing scattering, i.e., increasing $T_D$ moves the phase jump to higher magnetic fields.}
    \label{fig:phase_shifting}
\end{figure}

\end{document}